\documentclass[12pt]{article}
\usepackage{authblk}
\usepackage[bookmarksnumbered, colorlinks, plainpages]{hyperref}
\usepackage{amsmath, amsthm, amscd, amsfonts, amssymb, graphicx, color, booktabs}
\usepackage[left=2cm,right=2cm,top=2cm,bottom=2cm]{geometry}

\numberwithin{equation}{section}
\definecolor{email}{rgb}{0.00,0.00,0.84}

\usepackage{epsf,epsfig,psfrag,graphics,epstopdf}

\usepackage{float}

\usepackage{xcolor}

\usepackage{amsmath}
\usepackage{amsfonts}
\usepackage{amssymb}
\usepackage{graphicx}

\usepackage{cite}

\usepackage{slashed}

\usepackage{color}
\definecolor{nicered}{rgb}{0.7,0.1,0.1}
\definecolor{nicegreen}{rgb}{0.1,0.5,.1}

\usepackage{caption}
\usepackage{subcaption}

\usepackage{hyperref}
\hypersetup{colorlinks,citecolor=nicegreen,linkcolor=nicered}
\hypersetup{colorlinks=true}

\usepackage{comment}

\usepackage{rotating}

\usepackage{multirow}

\begin{document}
\setcounter{page}{1}




\title{\large \bf 12th Workshop on the CKM Unitarity Triangle\\ Santiago de Compostela, 18-22 September 2023 \\ \vspace{0.3cm}
\LARGE 2023 update of the extraction of the\\ CKM matrix elements}

\author{Luiz Vale Silva, on behalf of the \textsf{\textit{CKM}fitter} Collaboration \\
        Departament de F\'{i}sica Te\`{o}rica, Instituto de F\'{i}sica Corpuscular, \\ Universitat de Val\`encia -- Consejo Superior de Investigaciones Cient\'{i}ficas, \\ Parc Cient\'{i}fic, Catedr\'{a}tico Jos\'{e} Beltr\'{a}n 2, E-46980 Paterna, Valencia, Spain}

\maketitle

\begin{abstract}

I discuss the extraction of the Cabibbo-Kobayashi-Maskawa (CKM) matrix elements under the Standard Model (SM) framework from a global fit combining observables that satisfy the double requirement of being precisely known both experimentally and theoretically. The analysis shown here relies on the \textsf{\textit{CKM}fitter} package, consisting of a frequentist approach that employs the \textit{Range} fit (\textit{R}fit) scheme to handle theoretical uncertainties.

\end{abstract} \maketitle




\section{Introduction}



Flavor changing processes provide a powerful test of the SM, and consist, therefore, in an avenue to look for its extension.
The breaking of flavor symmetries in the SM originates from the Yukawa couplings of the SM Higgs, from which the spectrum of fermion masses and the CKM matrix in the quark sector emerge.
This matrix collects the fundamental parameters of the SM describing quark flavor mixing.
The CKM matrix in the SM is a three by three unitary matrix that can be parameterized by three mixing angles and one single source of CP violation \cite{Kobayashi:1973fv}.
Its introduction represents one of the most important achievements in fundamental particle physics, which celebrated recently its 50th anniversary. The year of 2023 held another important decennial anniversary, the one of the formulation of universality in $\Delta S = 0, 1$ weak transitions \cite{Cabibbo:1963yz}.
See Ref.~\cite{Olsen:2023lrt} for a historical note.
Having reached a mature age, much excitement is still expected ahead, particularly with the advent of LHCb Upgrades and Belle~II.

The CKM matrix turns out to be hierarchical, namely, elements closer to the diagonal are larger. A useful, rephasing invariant, parameterization that shows clearly this feature is the Wolfenstein parameterization consisting of the real parameters $ A $, $ \lambda $, $ \bar\rho $ and $ \bar\eta $

\begin{equation}
	{\color{black} \lambda} = \frac{\vert V_{us} \vert}{(\vert V_{ud} \vert^2 + \vert V_{us} \vert^2)^{1/2}} \, , \quad {\color{black} A \lambda^2} = \frac{\vert V_{cb} \vert}{(\vert V_{ud} \vert^2 + \vert V_{us} \vert^2)^{1/2}} \, , \quad {\color{black} \bar\rho+i \bar\eta} = - \frac{V^{}_{ud} V^{*}_{ub}}{V^{}_{cd} V^{*}_{cb}} \,,
\end{equation}
where no assumption has been made about the size of $ \lambda $, and a small $ \lambda $ accommodates the hierarchical structure mentioned above.
A useful graphical representation of the CKM matrix results from its unitarity: see the left panel of Fig.~\ref{fig:rhoetaBd_pulls} for the so-called $B_d$ unitarity triangle.

A rich variety of observables is used to probe the structure of the CKM matrix \cite{Charles:2004jd,Charles:2015gya,Koppenburg:2017mad}, including processes dominated by SM tree-level contributions, processes for which in the SM there is no tree-level contribution, processes that individually rule out a vanishing CP violating phase, or yet processes that are individually compatible with the absence of CP violation.
Some observables are presently dominated by experimental uncertainties (e.g., $\alpha$, $\beta$, $\gamma$), while others are dominated by systematic sources of uncertainty,
the latter being mainly due to hadronic effects inherent to quark flavor processes.
We heavily rely on Lattice QCD extractions of the hadronic inputs, e.g., bag parameters, decay constants and form factors.

Thanks to the experimental (e.g., $ B $-factories,\footnote{Our list of decennial anniversaries in 2023 closes with BaBar, that turned 30 years young since its approval.} LHCb, etc.) and theoretical (e.g., Lattice QCD) developments over the last two decades, an exquisite accuracy has been reached in the extraction of the CKM matrix elements. Here, we briefly discuss the 2023 update of the combination and extraction in the SM of the CKM matrix elements by the \textsf{\textit{CKM}fitter} Collaboration~\cite{Charles:2004jd}.
Previous results can be found in Ref.~\cite{Qian:2023eok}, that summarizes our 2021 update; see also Ref.~\cite{ValeSilva:2018nus} for the 2018 update.
Comparisons with the fits performed by the UTfit Collaboration \cite{Ciuchini:2000de,UTfit:2022hsi} can be found in \textsf{PDG} ``CKM Quark-Mixing Matrix'' mini-reviews.
Our primary goal is to test the SM, and possibly point out tensions in its description of flavor transitions, likely to occur in presence of physics beyond the SM.


\section{Statistical approach, overview of Lattice QCD inputs}


The \textsf{\textit{CKM}fitter} Collaboration employs a frequentist framework based on a $ \chi^2 $ analysis. The approach used to incorporate and manipulate systematic uncertainties is called \textit{Range} fit (\textit{R}fit) scheme. In practice, it means that one varies freely the true value of the fixed (and thus not stochastic) but unknown theoretical correction $ \delta $ inside the quoted uncertainty $ \pm \Delta $, i.e., $ \delta \in [- \Delta, \Delta] $, without any penalty from the $ \chi^2 $. This setup implies a plateau in the p-value of the input for
the parameter carrying the theoretical uncertainty $ \pm \Delta $ (see, e.g., Fig.~\ref{fig:Vub_Vcb_1D} below). Subsequently, Confidence Level (C.L.) intervals are determined by varying the $ \chi^2 $ around the best-fit point determined by the argument of $\chi^2_{min}$, which sets the goodness of the fit.


As previously stated,
important systematic, or theoretical, uncertainties are usually present in the quark flavor sector.
While here we adopt the \textit{R}fit scheme introduced above, other approaches are discussed and compared in Ref.~\cite{Charles:2016qtt}.
Though their treatment is somewhat ill-defined, a useful and meaningful modeling of systematic uncertainties must satisfy a certain number of criteria, such as good coverage properties (at least for the confidence level significances we are interested in),
the more technical requirement of being simple enough to allow for tractable calculations (i.e., the determination of the best-fit point and confidence intervals) when dealing with a large number of constraints and parameters to be extracted, etc. See, e.g., Ref.~\cite{Charles:2016qtt} for a more detailed discussion.




The Lattice QCD inputs used here, assessing hadronic quantities such as bag parameters, decay constants, form factors, etc., come from published references only (including proceedings), with uncertainty budgets provided, and consist of unquenched results with $ 2 $, $ 2+1 $, or $ 2+1+1 $ dynamical fermions.
We follow the \textsf{FLAG} review \cite{FlavourLatticeAveragingGroupFLAG:2021npn}, together with its reviewed version as of February 2023, in order to keep track of relevant Lattice QCD extractions, and do not consider results that carry red flags according to their criteria, except for $ \Lambda_b \to (\Lambda_c, p) $ form factors for which no alternative extraction is available.
If for a quantity we have many sources of systematic uncertainty, they are treated at the same footing and, in the absence of correlations, summed linearly (see Ref.~\cite{Charles:2016qtt} for a discussion of the definition and inclusion of correlations among theoretical uncertainties in the same spirit of the \textit{R}fit model for this category of uncertainties).
Different extractions of the same quantity are combined following a conservative procedure called \textit{educated Rfit} in which the averaged theoretical uncertainty is not smaller than the smallest of the individual theoretical uncertainties. More details are provided in Refs.~\cite{Charles:2004jd,Charles:2015gya,Charles:2016qtt}.
The averages in most cases are dominated by a single (i.e., $ f_K / f_\pi $, $f_{D_s}$, $f_{D_s}/f_D$, $ f_{B_s} $, $f_+^{D \to \pi} (0)$) or by two (i.e., $ f_K $, $f_+^{K \to \pi} (0)$, $f_+^{D \to K} (0)$, $ \hat{B}_K $, $ \hat{B}_{B_s} $, $ \hat{B}_{B_s} / \hat{B}_{B_d} $) extractions, and in some cases by three (i.e., $ f_{B_s}/f_B $) or more (i.e., $\bar{m}_c(\bar{m}_c)$).
Except for the bag parameters and $\bar{m}_c(\bar{m}_c)$, $N_f = 2+1+1$ extractions are the most precise, and except in the latter case the individual inputs are very consistent among themselves.







\section{Results and discussion}

\begin{figure}[t]
	\centering
    \includegraphics[scale=0.4,trim={0.5cm 0.5cm 0.3cm 1.8cm},clip]{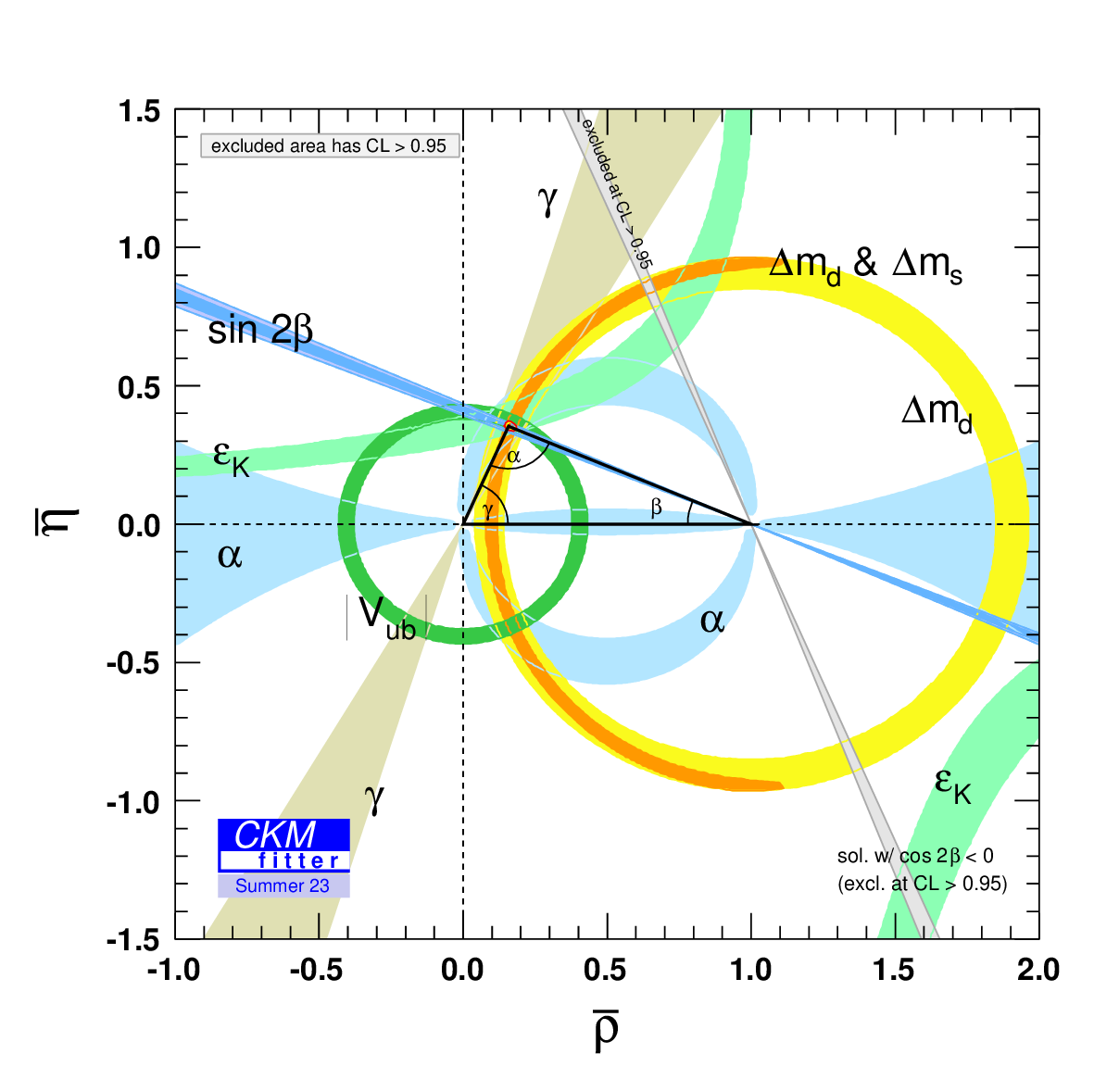} \hspace{8mm}
	\includegraphics[scale=0.415,trim={3cm 0.5cm 0.5cm 0.5cm},clip]{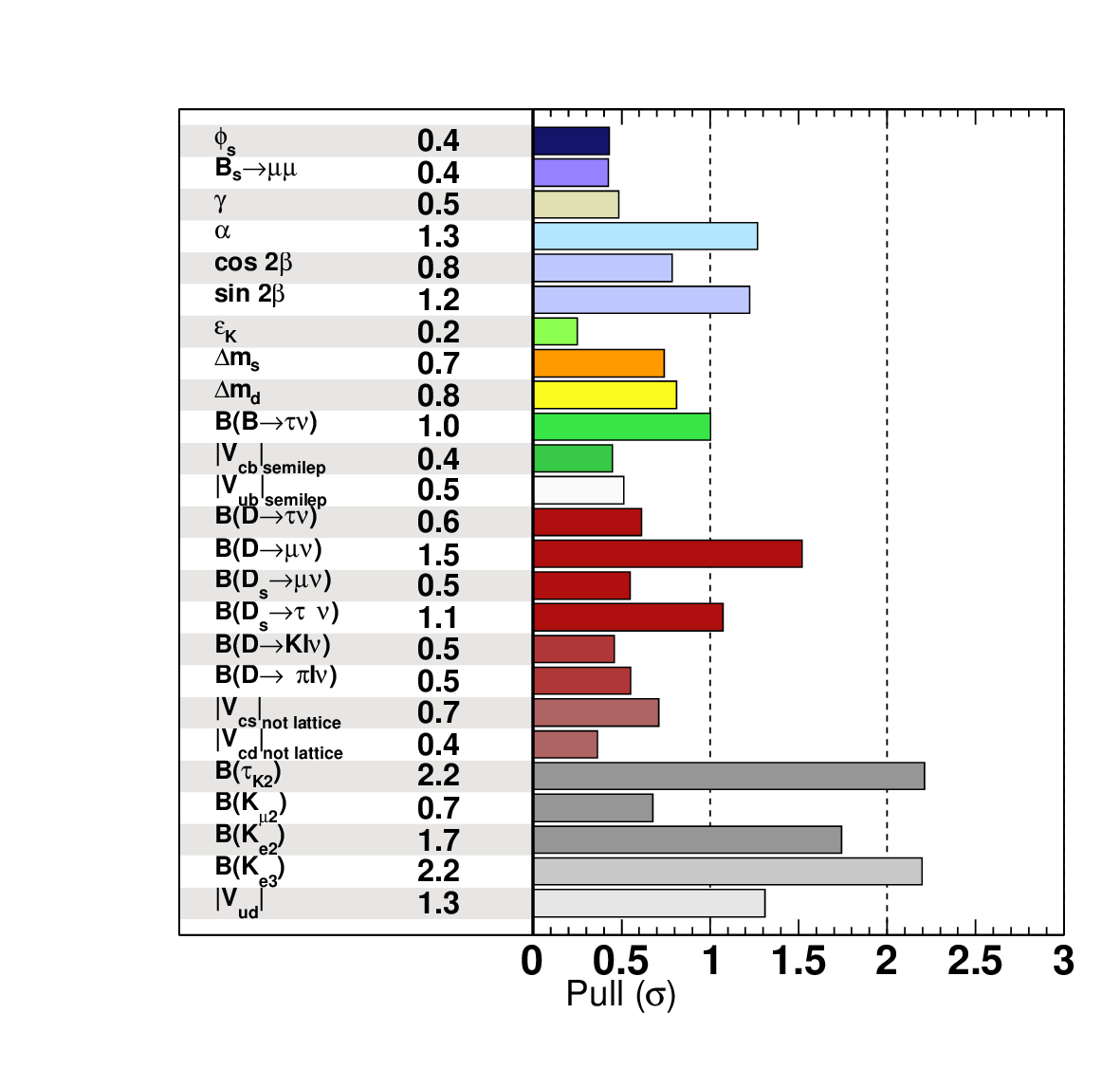}
	\caption{{\color{nicegreen} \it Left panel.} Combination of constraints used in the extraction of the Wolfenstein parameters in the $ (\bar\rho \,, \bar\eta) $ plane. The apex of the triangle, at $ \bar{\rho} \simeq 0.16 $ and $ \bar{\eta} \simeq 0.36 $, corresponds to the best-fit point. Also shown are the $ 68~\% $ (hashed red) and $ 95~\% $ (yellow with red contour) C.L. regions of the true value of $ \bar{\rho} $ and $ \bar{\eta} $. {\color{nicegreen} \it Right panel.} Pulls for individual observables in units of $ \sigma $.}\label{fig:rhoetaBd_pulls}
\end{figure}


The full set of experimental and theoretical inputs is given in Ref.~\cite{website}, summarized in Tab.~\ref{tab:expinputs}.
Compared to our Spring 2021 edition (following the conference Moriond~2021) \cite{Qian:2023eok}, the results of the global fit under the SM hypothesis remain excellent: the p-value is $ 67\% $, which corresponds to $ 0.4\sigma $, if all uncertainties are treated as Gaussian.
The consistent overall picture allows for a meaningful interpretation of flavor data in terms of the SM CKM matrix, the extracted Wolfenstein parameters being ($ 68\% $ C.L. intervals are quoted, except when explicitly stated otherwise)



\begin{eqnarray}\label{eq:fullGlobalFitExtraction}
A = 0.8215^{\,+0.0047}_{\,-0.0082} \, (0.8\% \, {\rm unc.})\,, &\qquad&
\lambda = 0.22498^{\,+0.00023}_{\,-0.00021} \, (0.1\% \, {\rm unc.})\,,\\
\bar\rho = 0.1562^{\,+0.0112}_{\,-0.0040} \, (4.9\% \, {\rm unc.})\,, &\qquad&
\bar\eta = 0.3551^{\,+0.0051}_{\,-0.0057} \, (1.5\% \, {\rm unc.})\,.\nonumber
\end{eqnarray}
In the global fit analysis, the small parameter $ \lambda $ is precisely determined from  $ | V_{ud} | $ (superallowed nuclear transitions) and $ | V_{us} | $ (semileptonic and leptonic kaon decays), while the parameter $ A $ is precisely determined from a combination of $ | V_{cb} | $, $ | V_{ub} | $, and further information from the global fit. The extraction of $ \bar{\rho} $ and $ \bar{\eta} $ is illustrated in the left panel of Fig.~\ref{fig:rhoetaBd_pulls}, where it is clear the dominant role played by $ \sin (2 \beta) $ and $ \Delta m_{d} / \Delta m_{s} $, but also $ | V_{ub} |/| V_{cb} | $, in the determination of the CP violating phase. For comparison,
the Wolfenstein parameters extracted from a global fit including only observables dominated by SM tree-level contributions, for which the p-value computed with Gaussian uncertainties is $ 40\%~(0.9\sigma) $, are 


\begin{eqnarray}
A = 0.8221^{\,+0.0050}_{\,-0.0189} \, (1\% \, {\rm unc.})\,, &\qquad&
\lambda = 0.22498^{\,+0.00022}_{\,-0.00021} \, (0.1\% \, {\rm unc.})\,,\\
\bar\rho = 0.151^{\,+0.016}_{\,-0.017} \, (11\% \, {\rm unc.})\,, &\qquad&
\bar\eta = 0.381^{\,+0.011}_{\,-0.023} \, (4\% \, {\rm unc.})\,,\nonumber
\end{eqnarray}
in good agreement with the extraction from the full global fit shown in Eq.~\eqref{eq:fullGlobalFitExtraction}.





\begin{figure}[t]
	\centering
    \includegraphics[scale=0.35,trim={0.5cm 0.5cm 0.3cm 0.5cm},clip]{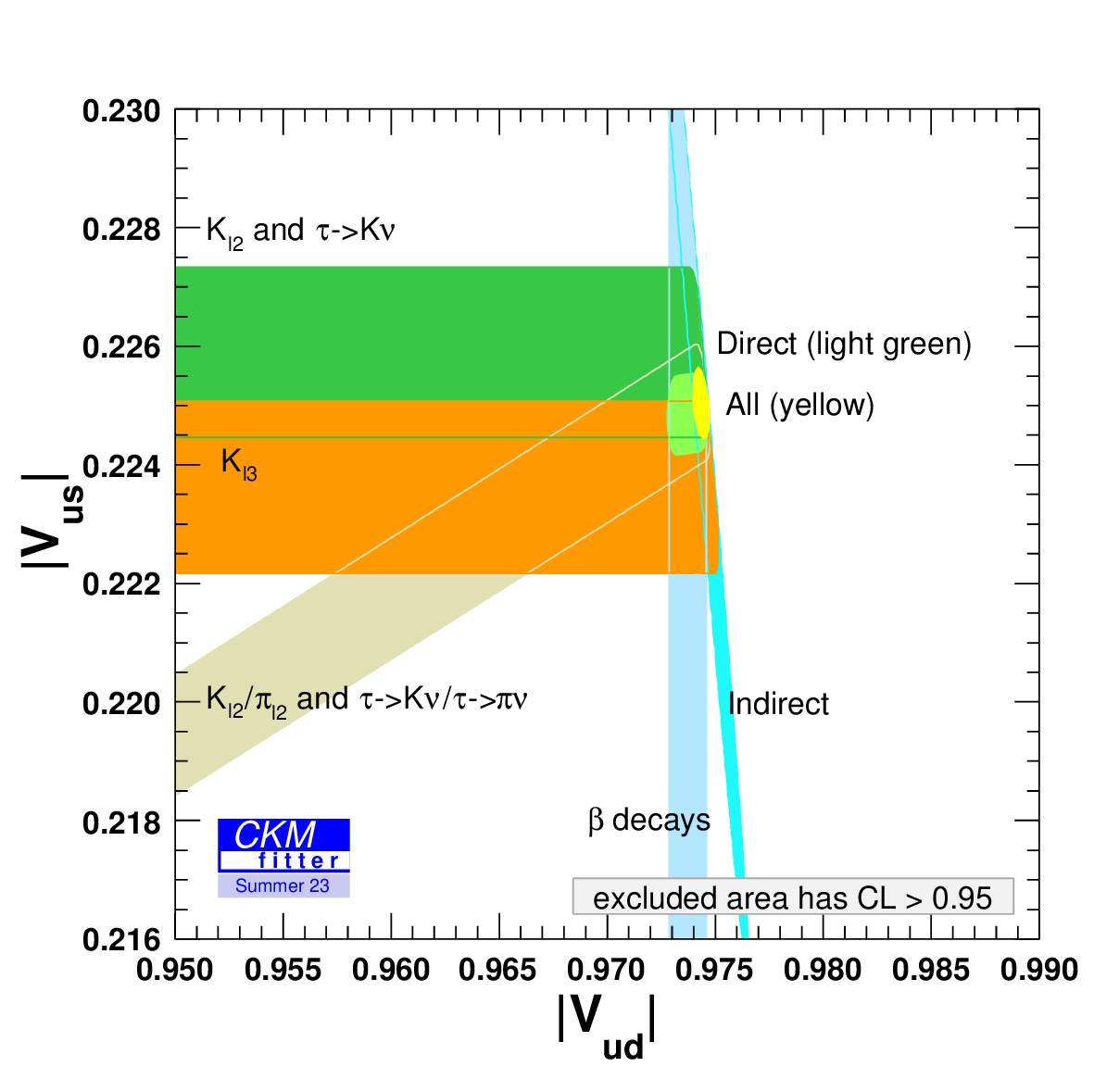} \hspace{8mm}
    \includegraphics[scale=0.4,trim={0.5cm 0.5cm 0.5cm 0.5cm},clip]{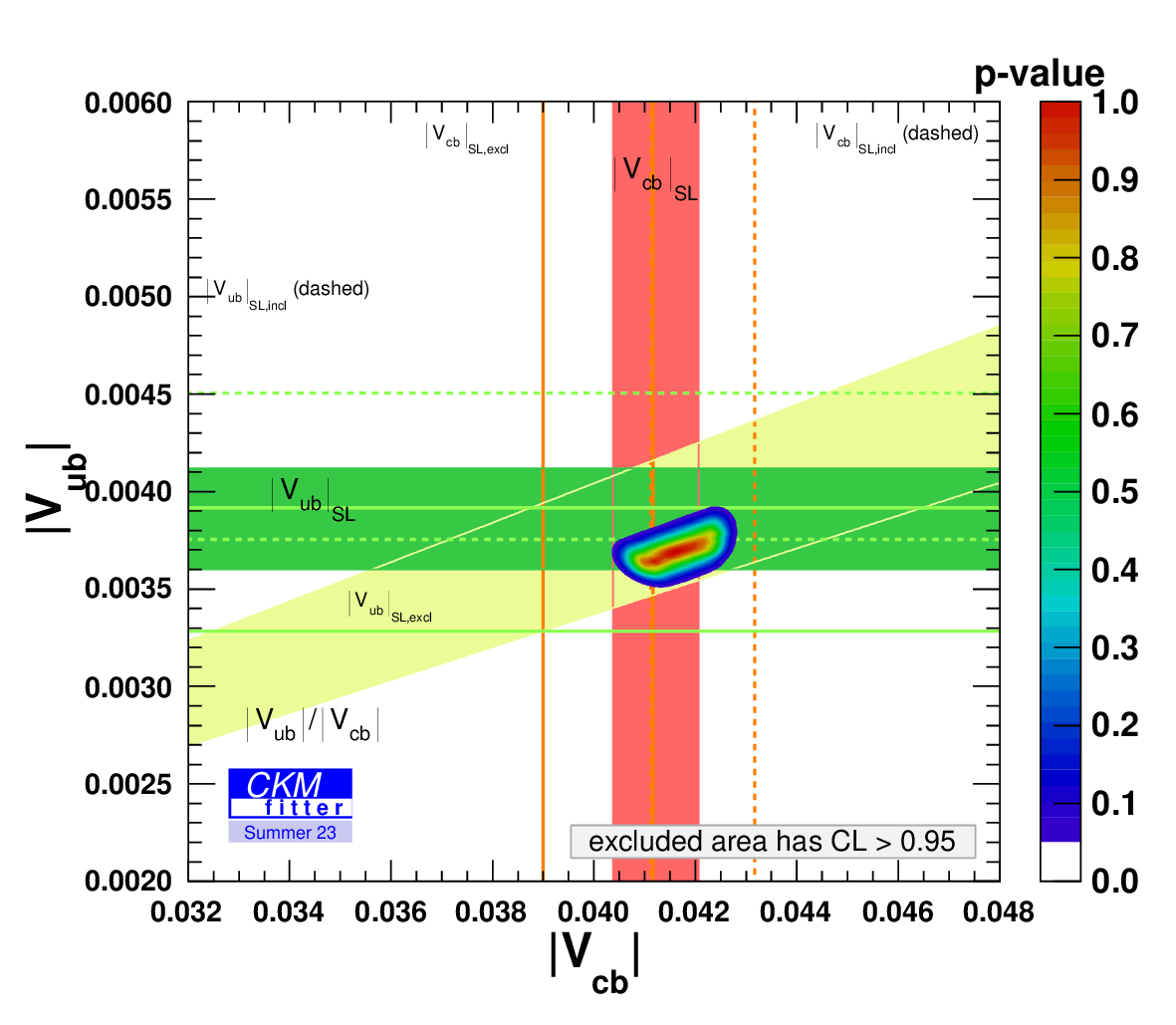}
	\caption{{\color{nicegreen} \it Left panel.} Status of the $ ( | V_{ud} | \, , | V_{us} | ) $ plane. The indirect constraints (coming from $b$ transitions) are related to $| V_{ud} |$ and $| V_{us} |$ through unitarity. The yellow region of the global combination corresponds to 68\% C.L. {\color{nicegreen} \it Right panel.} Status of the $ ( | V_{cb} | \, , | V_{ub} | ) $ plane. The horizontal and vertical coloured bands represent our averages of the determinations from semileptonic $ B $ decays. The white bands with solid (dashed) borders correspond to the determinations from exclusive (inclusive) semileptonic $ B $ decays. The diagonal coloured band corresponds to the determination of $ | V_{ub} |/| V_{cb} | $ from $ \Lambda_b $ decays, $B_s$-meson decays, and inclusively. The rainbow oval region indicates the indirect determination of $ | V_{ub} | $ and $ | V_{cb} | $ from the global fit, without any information from semileptonic or leptonic decays of $ b $-hadrons.}\label{fig:VudVus_VubVcb}
\end{figure}

Each observable considered in the fit provides a test of the validity of the SM, as measured by the pulls defined by

\begin{equation}
	Pull_{\mathcal{O}_{}} = \sqrt{\chi^2_{min} - \chi^2_{min, !\mathcal{O}_{}}} \,,
\end{equation}
where $ !\mathcal{O}_{} $ means that the corresponding $ \chi^2 $ is built without the input for $ \mathcal{O}_{} $.
The values of the pulls given in the right panel of Fig.~\ref{fig:rhoetaBd_pulls} imply that the SM predictions are in good agreement with the measurements. Since the observables have correlated fits (e.g., the rainbow oval region in the $ ( | V_{cb} | \, , | V_{ub} | ) $ plane shown in the right panel of Fig.~\ref{fig:VudVus_VubVcb}), and since theoretical uncertainties are not treated as Gaussian, the distribution of pulls is not normal, and the number of observables which have a pull larger than $ n \times \sigma $, for a certain $ n $, is not always a meaningful information.
Note that the presence of a plateau in the \textit{R}fit model for theoretical uncertainties may lead to a vanishing pull for some quantities even in cases where the predicted and the observed values are not identical.





\subsection{Discussion on some specific inputs}

Let us in turn discuss some of the inputs used in the fit, starting from the moduli of the CKM matrix elements.
Theoretical uncertainties are sizable in the extraction of $ |V_{ud}| $ \cite{Hardy:2020qwl},
motivating the revision of nuclear-structure dependent corrections. In agreement with the \textit{R}fit scheme, we combine theoretical uncertainties linearly. The resulting uncertainty is still smaller than the one coming from neutron $\beta$ decays by a factor $\lesssim 2$.
In testing first row unitarity,
isospin breaking corrections in $P_{\ell 2}$, $P=K, \pi$, decays are currently taken from Ref.~\cite{Cirigliano:2011tm}.
The left panel of Fig.~\ref{fig:VudVus_VubVcb} displays the different inputs used in the extractions of $| V_{ux} |$, $x=d, s$.
Deviation from unitarity is not significant,
$ ( | V_{ud} |^2 + | V_{us} |^2 + | V_{ub} |^2 - 1 ) \times 10^3 = (-2.54, -0.12) $ [$1\sigma$], and $(-2.75, 0.02)$ [$2\sigma$].

There has been significant progress in the direct extractions of $ | V_{cd} | $ and $ | V_{cs} | $, led from both better theoretical inputs for the form factors \cite{Chakraborty:2021qav,FermilabLattice:2022gku}, and experimental results, see, e.g., Ref.~\cite{BESIII:2023fhe} and references therein. However, their extractions are still dominated by indirect determinations.
Comparisons of the impact of different categories of inputs are found in Ref.~\cite{website}. Second row unitarity is well respected, namely, $( | V_{cd} |^2 + | V_{cs} |^2 + | V_{cb} |^2 - 1 ) \times 10^3 = (-1.1, 1.7)$ [$1\sigma$].

\begin{figure}[t]
	\centering
    \includegraphics[scale=0.33,trim={0.5cm 0.4cm 0.3cm 0.5cm},clip]{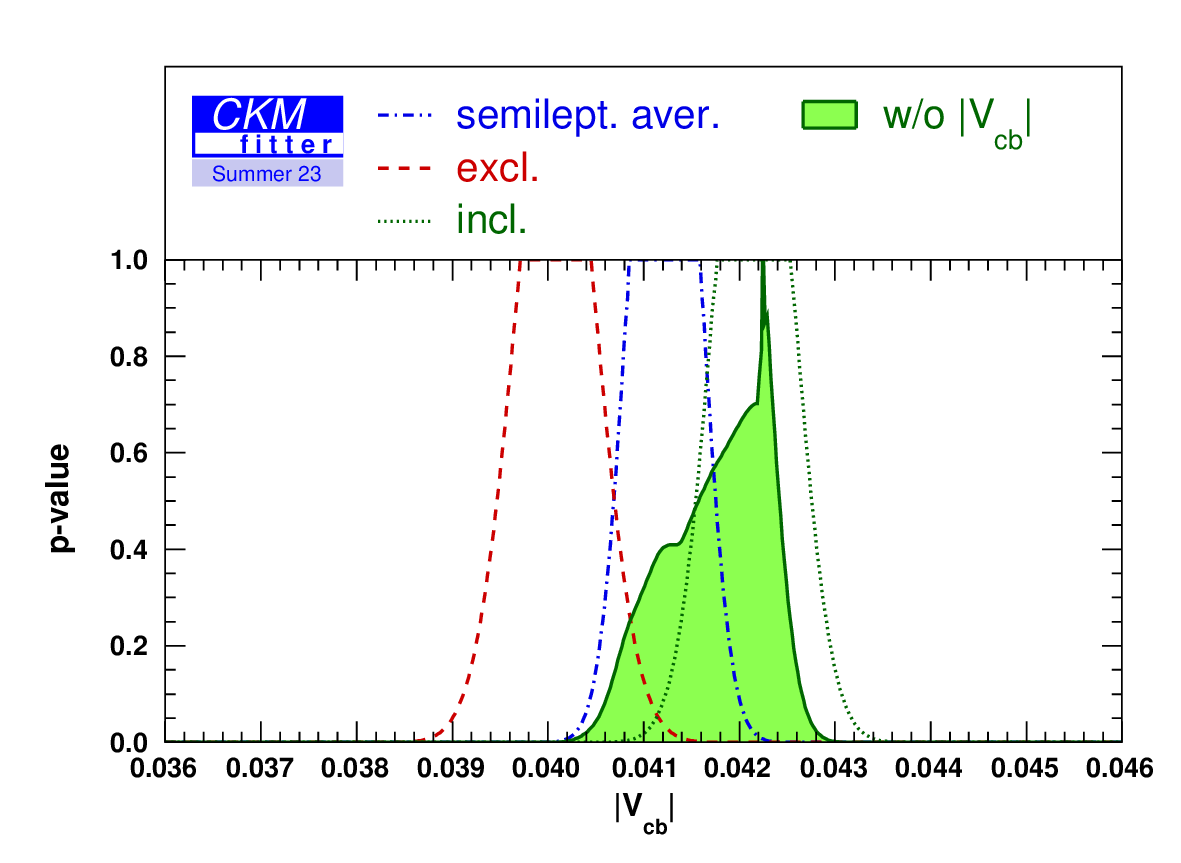} \hspace{10mm}
    \includegraphics[scale=0.33,trim={0.5cm 0.4cm 0.3cm 0.5cm},clip]{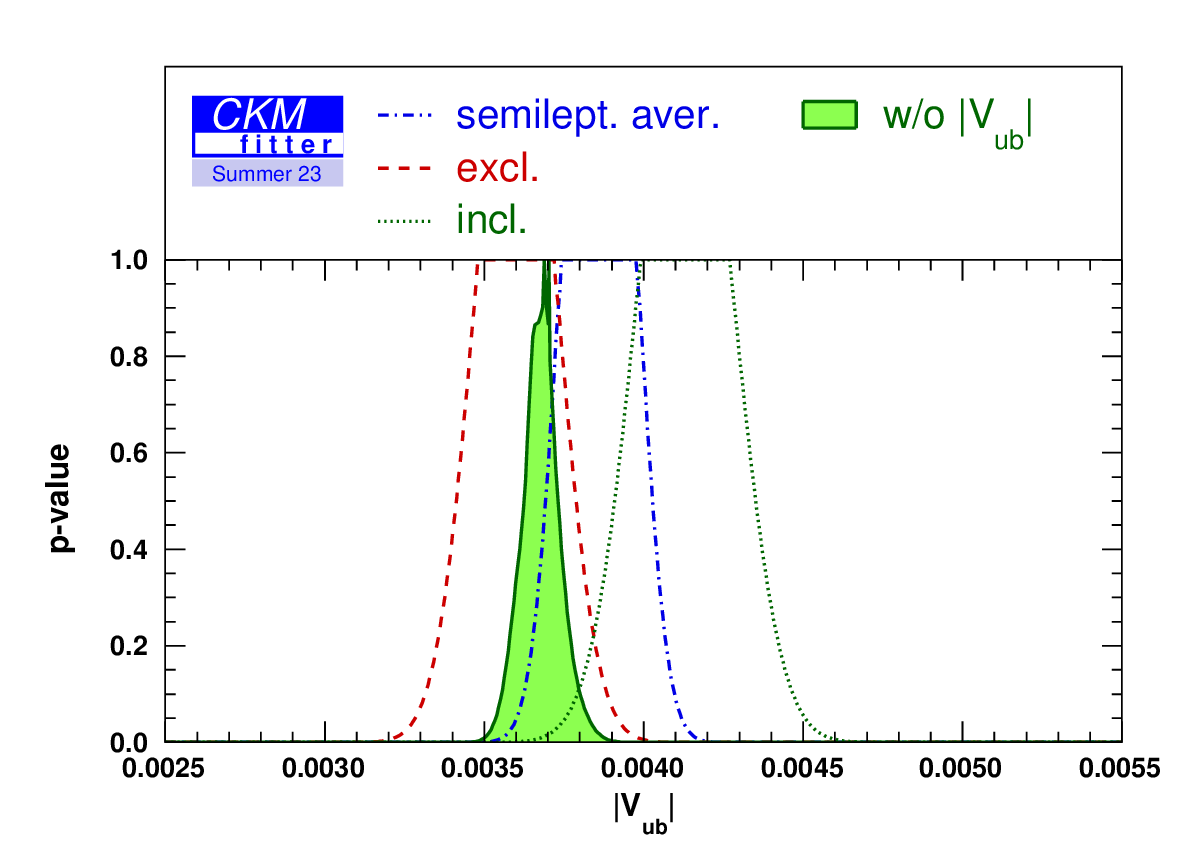}
	\caption{Constraints on $| V_{cb} |$ (left) and $| V_{ub} |$ (right) from inclusive and exclusive $B$ decays, as well as our averages, compared to the indirect determinations from the global fit.}\label{fig:Vub_Vcb_1D}
\end{figure}

We now discuss semileptonic extractions of $ | V_{xb} | $, $x=u,c$, for which there have been long-standing tensions between inclusive and exclusive extractions.
To start, the exclusive extraction $|V_{cb}|_{excl.}=(40.08\pm0.36\pm0.37)\times 10^{-3}$ combines values from $ B \to D \ell \nu_\ell $ (BCL) \cite{HFLAV:2022esi,FlavourLatticeAveragingGroupFLAG:2021npn} and $ B \to D^\ast \ell \nu_\ell $ (BGL only).
The latter includes new results from Belle (tagged) \cite{Belle:2023xgj} and Belle II (untagged) \cite{Belle-II:2023okj}.
Lattice QCD inputs for the $ B \to D^\ast \ell \nu_\ell $ form factors come from Fermilab-MILC \cite{FermilabLattice:2021cdg}, HPQCD \cite{Harrison:2023dzh}, and JLQCD \cite{Aoki:2023qpa};
see the newest \textsf{PDG} ``Semileptonic $b$-Hadron Decays, Determination of $V_{cb}$, $V_{ub}$'' mini-review for further discussion.
For the inclusive extraction, we have considered $ |V_{cb}|_{incl.}=( 42.15\pm0.32\pm0.39 ) \times 10^{-3} $ \cite{Bordone:2021oof} (that employs $m_b^{kin}$).
Note that the inputs for exclusive and inclusive values have similar error budgets, although coming from very different experimental and theoretical methods, and differ by about $2\sigma$; see the left panel of Fig.~\ref{fig:Vub_Vcb_1D}. Their combination using the educated \textit{R}fit approach leads to the following average


\begin{equation}
	|V_{cb}|_{\rm SL}=( 41.22 \pm 0.24 \pm 0.37 ) \times 10^{-3} \,,
\end{equation}
which is similar to our 2019 value (namely, $|V_{cb}|_{\rm SL}=( 41.15 \pm 0.34 \pm 0.45 ) \times 10^{-3}$). The indirect determination relative uncertainty based on CKM unitarity (and information on the ratio $|V_{ub} / V_{cb}|$ discussed below) amounts to $1.7\%$, while when using all the available information the relative uncertainty drops to $0.9\%$.

Concerning the inclusive $|V_{ub}|$,
we use the same inputs as in Ref.~\cite{HFLAV:2022esi} (see also the 2019 edition),
that includes the new measurements of partial branching fractions of $B \to X_u \ell \nu_\ell$~\cite{Belle:2021eni},
leading to $|V_{ub}|_{incl.}=(4.13 \pm 0.12 \pm 0.14)\times 10^{-3}$ (GGOU+BLNP+DGE).
For the exclusive extraction, we consider Ref.~\cite{FlavourLatticeAveragingGroupFLAG:2021npn},
together with the new, preliminary measurement by Belle II from $ B \to \pi \ell \nu_\ell $, 
resulting in $|V_{ub}|_{excl.}=(3.60 \pm 0.10 \pm 0.12)\times 10^{-3}$.
Again, the inputs for exclusive and inclusive values have similar error budgets, and are compatible at about $1\sigma$; see the right panel of Fig.~\ref{fig:Vub_Vcb_1D}.
Our average of the two values is obtained using the educated \textit{R}fit approach, resulting in

\begin{equation}
	|V_{ub}|_{\rm SL}=( 3.86 \pm 0.07 \pm 0.12 ) \times 10^{-3} \,,
\end{equation}
which is again similar to our 2019 value (namely, $|V_{ub}|_{\rm SL}=( 3.88 \pm 0.08 \pm 0.21 ) \times 10^{-3}$). The indirect extraction of $|V_{ub}|$ from the leptonic decay and CKM unitarity (excluding $|V_{ub} / V_{cb}|$ discussed below) carries a relative uncertainty of $2.1\%$, while from the full fit the relative uncertainty drops to $1.2\%$.

There are also distinct direct determinations of $ | V_{ub} / V_{cb} | $, from $ \Lambda_b $ decays \cite{LHCb:2015eia}
(updated according to the new $\Lambda_c^+ \to p K^- \pi^+$ branching ratio based on \textsf{PDG}), $B_s$-meson decays \cite{LHCb:2020cyw}
(using the high $q^2$ bin, where Lattice QCD inputs are available), and inclusively \cite{Belle:2023asa} (tagged, GGOU).
The latter is the most recent public result used in this update.
The status of semileptonic $ | V_{xb} | $, $x=u,c$, extractions is shown in the right panel of Fig.~\ref{fig:VudVus_VubVcb}.
As in previous editions,
the indirect fit still favors the inclusive (exclusive) semileptonic extraction of $ | V_{cb} | $ (respectively, $ | V_{ub} | $).


We now shift our attention to the extractions of the unitarity triangle angles.
We refer the reader to Ref.~\cite{Charles:2017evz} for a thorough discussion dedicated to the angle $\alpha$ ($\phi_2$). Compared to our previous edition, we include results from Belle II on $ B^+ \to \rho^+ \rho^0 $ \cite{Belle-II:2022ksf}, $ B^0 \to \rho^+ \rho^- $ \cite{Belle-II:2022ihd}, $ B^0 \to \pi^0 \pi^0 $ \cite{Belle-II:2023cbc}, $ B^+ \to \pi^+ \pi^0 $ \cite{Belle-II:2022mib}. The directly extracted value is $ \alpha = (86.2^{+3.9}_{-3.5})^{\rm o} $ (not displaying the solution around $ \alpha = 0 $), whose improvement with respect to the latest update (namely, $ \alpha = (86.4^{+4.3}_{-4.0})^{\rm o} $) comes essentially from the $ \pi \pi $ decay mode.

There has been an important progress in the extraction of the angle $\beta$ ($\phi_1$), with a substantial decrease of the uncertainty that reflects in a more precise extraction of the $(\bar\rho , \, \bar\eta)$ apex. We employ the value $ \sin 2 \beta = 0.708 \pm 0.011 $ from a preliminary \textsf{HFLAV} combination \cite{Li_Jevtic}, to be compared with our previous input $ \sin 2 \beta = 0.699 \pm 0.017 $. The analogous angle $ \beta_s $ has also been significantly improved, see Ref.~\cite{Li_Jevtic}.

Finally, we employ $\gamma$ ($\phi_3$) from experimental inputs up to December 2021 $ \gamma = (65.9^{+3.3}_{-3.5})^{\rm o} $ \cite{HFLAV:2022esi}. Note that this is a more precise value than the direct input used for the angle $\alpha$.

\section{Conclusions}

The SM gives a consistent picture of flavor transitions and CP violation in the quark sector: at the moment, the global fit does not point to large deviations from the SM. See Ref.~\cite{Charles:2020dfl} for future prospects concerning searches for NP in $ |\Delta F| = 2 $ processes. More information relative to the 2023 edition discussed here is found is our website, Ref.~\cite{website}.
The reader is invited to use the web interface Ref.~\cite{ckmlive} to prepare their own global fit analyses.


\vspace{0.5cm}


\noindent
\textbf{Acknowledgements.}
The results here discussed have been achieved in collaboration with all members of the \textsf{\textit{CKM}fitter} group.
This project has received funding from the European Union’s Horizon 2020 research and innovation programme under the Marie Sklodowska-Curie grant agreement No 101031558.
This work has been supported by MCIN/AEI/10.13039/501100011033, grant PID2020-114473GB-I00, and by Generalitat Valenciana, grant PROMETEO/2021/071.
I am grateful for the hospitality of the CERN-TH group where part of this research was executed.


\begin{sidewaystable}
\centering
\renewcommand\arraystretch{1.2}
{\footnotesize \begin{tabular}{c|c|cccc|ccc}
CKM  & Process  & \multicolumn{4}{c|}{Observables}  & \multicolumn{3}{c}{Non-perturbative theoretical inputs}\\
\hline
$|V_{ud}|$ & $0^+\to 0^+$ $\beta$
                  & $|V_{ud}|_{\rm nucl}$ &=& $0.97373 \pm 0.00009 \pm 0.00053$
                  & \cite{Hardy:2020qwl} & \multicolumn{3}{c}{Nuclear matrix elements} \\
                    \hline
\multirow{4}{*}{$|V_{us}|$} & $K\to\pi\ell\nu_\ell$ 
                  & $|V_{us}|_{\rm SL}f_+^{K\to\pi}(0)$ &=& $ 0.21635 \pm 0.00038 $ & \cite{Moulson_Passemar}
                  & $f_+^{K\to\pi}(0)$ &=& $0.9675 \pm 0.0011 \pm 0.0023$\\
                 &  $K\to e\nu_e$ 
                 & ${\cal B}(K\to e\nu_e)$&=&$(1.582 \pm 0.007)\cdot 10^{-5}$ & \cite{ParticleDataGroup:2022pth}
                 &  \multirow{3}{*}{$f_K$}&\multirow{3}{*}{=}& \multirow{3}{*}{$155.57\pm0.17\pm0.57 $ MeV} \\
                &  $K\to \mu\nu_\mu$ 
                &  ${\cal B}(K\to \mu\nu_\mu)$&=& $0.6356 \pm 0.0011$
                & \cite{ParticleDataGroup:2022pth}\\
                 &  $\tau \to K \nu_\tau$ 
                 & ${\cal B}(\tau \to K\nu_\tau)$&=&$(0.6986 \pm 0.0085)\cdot 10^{-2}$
                 & \cite{HFLAV:2022esi}\\
                 \hline
\multirow{2}{*}{$|V_{us}/V_{ud}|$}                 &  $K\to \mu\nu_\mu/\pi\to\mu\nu_\mu$ & 
                 $\frac{{\cal B}(K\to \mu\nu_\mu)}{{\cal B}(\pi \to \mu\nu_\mu)}$
                         &=&$1.3367 \pm 0.0028$
                 & \cite{ParticleDataGroup:2022pth} &
                 \multirow{2}{*}{$f_K/f_\pi$}&\multirow{2}{*}{=}&\multirow{2}{*}{$1.1973 \pm 0.0007\pm0.0014$}
                  \\
                 &  $\tau\to K\nu_\tau/\tau \to \pi\nu_\tau$ &   
                 $\frac{{\cal B}(\tau \to K\nu_\tau)}{{\cal B}(\tau \to \pi\nu_\tau)}$
                        &=& $(6.437 \pm 0.092)\cdot 10^{-2}$
                 & \cite{HFLAV:2022esi} \\
                 \hline
\multirow{4}{*}{$|V_{cd}|$}   & $\nu N$ & $|V_{cd}|_{\rm not \; lattice}$ &=& $0.230\pm 0.011$ & \cite{ParticleDataGroup:2022pth}\\
                   & $D\to \tau\nu_\tau $ & ${\cal B}(D\to \tau\nu_\tau)$ &=& $(1.20\pm0.27)\cdot 10^{-3}$ 
                   & \cite{HFLAV:2022esi}
                   &\multirow{2}{*}{$f_{D_s}/f_D$}&\multirow{2}{*}{=}&\multirow{2}{*}{$1.1782 \pm 0.0006\pm0.0033$}\\
                   & $D\to \mu\nu_\mu $ & ${\cal B}(D\to \mu\nu_\mu)$ &=& $(3.77\pm0.17)\cdot 10^{-4}$ 
                   & \cite{HFLAV:2022esi} & & & \\
                   & $D\to \pi\ell\nu_\ell $ & $|V_{cd}|_{\rm SL} f_+^{D\to \pi}(0)$ &=& $0.1426 \pm 0.0018$ 
                   & \cite{HFLAV:2022esi}
                   &$f_+^{D\to \pi}(0)$&=&$0.624\pm 0.004\pm 0.006 $\\
                   \hline
\multirow{4}{*}{$|V_{cs}|$}   &  $W\to c\bar{s}$ &   $|V_{cs}|_{\rm not \; lattice}$ &=& $0.967 \pm 0.011$ & \cite{CMS:2022mhs}\\ 
                   & $D_s\to \tau\nu_\tau$ 
                   & ${\cal B}(D_s\to \tau\nu_\tau)$&=& $(5.32 \pm 0.10) \cdot 10^{-2}$ 
                   &  \cite{BESIII:2023fhe}
                   & \multirow{2}{*}{$f_{D_s}$} &\multirow{2}{*}{=}& \multirow{2}{*}{$249.23\pm 0.27 \pm 0.65$ MeV}\\
                   & $D_s\to \mu\nu_\mu$ 
                   & ${\cal B}(D_s\to \mu\nu_\mu)$&=& $(5.43\pm0.16)\cdot 10^{-3}$ 
                   &  \cite{HFLAV:2022esi}\\
                   & $D\to K\ell\nu_\ell $ & $|V_{cs}|_{\rm SL} f_+^{D\to K}(0)$ &=& $0.7180 \pm 0.0033$ 
                   & \cite{HFLAV:2022esi}
                   &$f_+^{D\to K}(0)$&=&$ 0.742\pm0.002\pm0.004$\\
                   \hline
\multirow{2}{*}{$|V_{ub}|$} & semileptonic $B$
                  & $|V_{ub}|_{\rm SL}$ &=& $(3.86 \pm 0.07 \pm 0.12)\cdot 10^{-3}$ 
                  & see text &
                     \multicolumn{3}{c}{form factors, shape functions}\\
                  & $B\to \tau\nu_\tau$ 
                  & ${\cal B}(B\to\tau\nu_\tau)$ &=& $(1.09\pm0.24) \cdot 10^{-4}$ & \cite{HFLAV:2022esi}
                  &   $f_{B_s}/f_B$&=& $1.2118\pm 0.0020 \pm 0.0058 $\\
                  \hline
$|V_{cb}|$ & semileptonic $B$
                 & $|V_{cb}|_{\rm SL}$ &=& $(41.22\pm 0.24 \pm 0.37)\cdot 10^{-3}$ & see text
                 &  \multicolumn{3}{c}{form factors, OPE matrix elements}\\               
\hline
\multirow{3}{*}{$|V_{ub}/V_{cb}|$} & semileptonic $\Lambda_b$
                 & $\frac{\mathcal{B}(\Lambda_b\to p\mu\nu_\mu)_{q^2>15}}{\mathcal{B}(\Lambda_b\to \Lambda_c\mu\nu_\mu)_{q^2>7}}$ &=& $(0.918 \pm 0.083 ) \cdot 10^{-2}$ & \cite{LHCb:2015eia}
                 &  \multicolumn{3}{c}{$\frac{\zeta(\Lambda_b\to p\mu\nu_\mu)_{q^2>15}}{\zeta(\Lambda_b\to \Lambda_c\mu\nu_\mu)_{q^2>7}}=1.471\pm 0.096\pm 0.290$}
                 \\
                 & semileptonic $B_s$
                 & $\frac{\mathcal{B}(B_s \to K \mu\nu_\mu)_{q^2>7}}{\mathcal{B}(B_s \to D_s \mu\nu_\mu)}$ &=& $(3.25 \pm 0.28) \cdot 10^{-3}$ & \cite{LHCb:2020ist}
                 & \multicolumn{3}{c}{$\frac{\zeta(B_s \to K \mu\nu_\mu)_{q^2>7}}{\zeta(B_s \to D_s \mu\nu_\mu)}=0.363\pm 0\pm 0.065$}
                 \\
                 & inclusive
                 & $|V_{ub}/V_{cb}|_{\rm incl}$&=& $ 0.100 \pm 0.006 \pm 0.003 $ & \cite{Belle:2023asa}
                 & & & \\
\hline
$\alpha$ & $B\to\pi\pi$, $\rho\pi$, $\rho\rho$ 
                & \multicolumn{3}{c}{branching ratios, $CP$ asymmetries} & see text
                & \multicolumn{3}{c}{isospin symmetry}\\
                \hline
\multirow{1}{*}{$\beta$}   & $B\to (c\bar{c}) K$ 
               & $\sin(2\beta)_{[c\bar{c}]}$ &=& $0.708 \pm 0.011$ 
               & \cite{HFLAV:2022esi} & \multicolumn{3}{c}{subleading penguins neglected}\\
\hline
$\gamma$ & $B\to D^{(*)} K^{(*)}$ 
                 & $ \gamma $ &=& $ (65.9^{+3.3}_{-3.5})^{\rm o} $
                 &  \cite{HFLAV:2022esi}& \multicolumn{3}{c}{GGSZ, GLW, ADS methods} \\
                   \hline
$\phi_s$ & $B_s\to J/\psi (KK, \pi\pi)$ & $(\phi_s)_{b \rightarrow c \bar{c} s}$ &=& $-0.039\pm 0.016$
             & \cite{HFLAV:2022esi}&
 \\
  \hline  
\multirow{3}{*}{$V_{tq}^*V_{tb}$}       & $\Delta m_d$ 
                & $\Delta m_d$ &=& $0.5065 \pm 0.0019$ ps${}^{-1}$
                & \cite{HFLAV:2022esi}
                &  $\hat{B}_{B_s}/\hat{B}_{B_d}$ &=& $1.007 \pm 0.010\pm0.014$\\ 
                 & $\Delta m_s$ & $\Delta m_s$ &=& $17.765\pm0.006$ ps${}^{-1}$ 
                 & \cite{LHCb:2021moh}
                 & $\hat{B}_{B_s}$&=& $1.313\pm0.012\pm0.030$\\
                                 & $B_s\to \mu\mu$ & ${\cal B}(B_s\to\mu\mu)$ &=& $(3.45 \pm 0.29) \cdot 10^{-9} [\times (1 - 0.063)]$
                 & \cite{HFLAV:2022esi} & $f_{B_s}$ &=& $228.75\pm0.69\pm1.87 $ MeV \\
                 \hline
$V_{td}^*V_{ts}$ and  
      & $\varepsilon_K$ & $|\varepsilon_K|$ &=& $(2.228\pm0.011)\cdot 10^{-3}$
       & \cite{ParticleDataGroup:2022pth}
       &$\hat{B}_K$&=& $0.7567\pm0.0020\pm0.0123$\\
$V_{cd}^*V_{cs}$      &   &&&&& $\kappa_\varepsilon$&=& $0.940\pm0.013\pm0.023 $\\
\end{tabular}}
\caption{\it \footnotesize Constraints used for the global fit, and the main inputs involved (more information can be found in Ref.~\cite{website}). When two errors are quoted, the first one is statistical, and the second one systematic. In the cases of $ \alpha $ angle, many different channels or methods are used to extract their values, and the full resulting p-value profiles are used as the inputs for the global fit \cite{website}.
Perturbative inputs are used for $\Delta m_d$, $\Delta m_s$, $\varepsilon_K$ (short-distance QCD), ${\cal B}(K\to \mu\nu_\mu) / {\cal B}(\pi \to \mu\nu_\mu)$, ${\cal B}(\tau \to K\nu_\tau) / {\cal B}(\tau \to \pi\nu_\tau)$ (EM corrections).
The remaining inputs are the masses of the charm and top quarks, and the value of the strong coupling.
\label{tab:expinputs}}
\end{sidewaystable}

\bibliography{mybib}{}
\bibliographystyle{unsrturl}

\end{document}